# Orchestrating Mixed-Criticality Cloud Workloads in Reconfigurable Manufacturing Systems

Marco Barletta, Marcello Cinque, Davide De Vita
University of Naples Federico II Via Claudio 21, 80125 Naples, Italy
macinque@unina.it

*Abstract*—The adoption of cloud computing technologies in the industry is paving the way to new manufacturing paradigms. In this paper we propose a model to optimize the orchestration of workloads with differentiated criticality levels on a cloud-enabled factory floor. Preliminary results show that it is possible to optimize the guarantees to deployed jobs without penalizing the number of schedulable jobs. We indicate future research paths to quantitatively evaluate job isolation.

## I. Introduction

The ongoing digitalization of industries, following the Industry 4.0 (I4.0) revolution is pushing the use of cloud technologies as a means to create Reconfigurable Manufacturing Systems (RMSs) that support the rapid addition, removal, or modification of process controls, functions, and/or operations, through reconfigurable hardware and software, to scale production capability and capacity [1].

The use of cloud and virtualization technologies in industrial settings [2] allows a simplification of the RMS management, adopting well consolidated practices on software orchestration and DevOps. This enables the use of commodity and cheap hardware on the factory floor or the consolidation of multiple functionalities, implemented as virtual machines or containers on heterogeneous machines. This meets SWaP (Size, Weight and Power) requirements, enabling the management of scalability and failovers with the same approaches used for cloud services, such as replication and migration.

In such settings, when the factory floor needs to be re-purposed for different productions, (virtual) Programmable Logic Controllers can be flexibly deployed side-by-side with a supervisory system (e.g., a SCADA) and a management GUI, even all on the same machine, disrupting the strict layered architecture mandated by the Computer Integrated Manufacturing (CIM) pyramid [3]. However, this type of flexibility requires to run workloads with differentiated real-time and safety requirements on the same shared hardware, which is typical of Mixed-Criticality Systems [4].

To meet such requirements, cloud orchestration platforms (e.g, Kubernetes, Docker Swarm) must be extended [5]–[7] with the notion of workload *criticality* and *assurance* level of computing nodes, i.e., the degree of isolation and dependability that a node can provide based on its hardware/software characteristics and current load. It is desirable to deploy highly critical workloads on high-assurance nodes, capable of isolating their execution from interferences through techniques like partitioning hypervisors and/or real-time CPUs.

In this paper we complement the orchestration model presented in our previous work, named K4.0s [6], with a scheduling model and an algorithm to optimally place mixed-criticality workloads on nodes with different assurance levels. The scheduling problem is modeled as a multi-objective problem that accounts for job acceptance, total assurance, and free resources. Preliminary results show, in a simulated environment, the benefits of accounting for the assurance levels during the scheduling. Finally, we indicate future research directions to quantitatively compute and predict the assurance levels through causal reasoning and Bayesian networks, which are a good fit for the proposed scheduler.

## II. Orchestration model

We adopt the model defined in [6], simplifying it to use only abstractions required in this paper. Hence, we model a Worker Node as $WN_i = <\vec{BR}_i, A_i(t), RT_i, Jobs_i> \; i \in \mathbf{N}$, where $\vec{BR}_i$ is a vector of *basic hardware resources total capacities* (CPU, Disk and Memory). $RT_i \in \{RT, nonRT\}$ is the node's *real-time capability*. $Jobs_i$ is the set of jobs assigned to the node $i$. $A_i(t) = f(\alpha_i, \beta_i, \gamma_i(t)) \in [0,1]$ is the *assurance level*, where $\alpha_i$ and $\beta_i$ accounts for, respectively, hardware and software; while $\gamma_i(t)$ is inversely related to the load of the node.

We model a job as $J_j = <\vec{BR}_j, C_j, RT_j> \; j \in \mathbf{N}$, where $RT_j$ is the same as defined above; $\vec{BR}_j$ accounts for the minimum (required) and maximum (to handle overloads) amount of resources to be allotted to a job (respectively $br_{request}$ and $br_{limit}$, $\forall_{br \in \vec{BR}_j}$); $C_j \in \{NO, LOW, HIGH\}$ is the *criticality level*. We define the $\theta_{LOW}$ and $\theta_{HIGH}$ thresholds of minimum assurance required at anytime to keep hosting critical jobs.

### A. Scheduling optimization problem

We model the scheduling problem as a multi-objective problem. Our objective function is a weighted sum of the following objective functions to maximize, each of them designed such that the codomain is [0; 1]:

- **Acceptance rate**: percentage of jobs that the scheduler managed to assign to its nodes.
  $\frac{\sum_i |Jobs_i|}{m}$, with $m$ number of pending + active jobs
  This function forces the scheduler to assign jobs instead of rejecting them.
- **Node assurances**: current cluster average assurance
  $\sum_{i:\,RT_i=RT} A_i(t) \Big/ \sum_{i:\,RT_i=RT} 1$
  The function accounts for the assurance loss that the jobs may cause to already loaded nodes, and to ensure that incoming critical jobs find suitable nodes to host them. The assurance value is a percentage of $\alpha + \beta$, which increases in thresholds defined by the minimum job criticality level that has guaranteed resources and whether the guaranteed resources are minimum or maximum.

- **Residual capacity**: sum of nodes' squared free resources

$$\sum_i \sum_{z=1}^{|\vec{BR}_i|} (\vec{BR}_i(z) - \sum_{j \in Jobs_i} \vec{BR}_j(z)_{limit})^2 \Big/ \sum_{z=1}^{|\vec{BR}_i|} (\vec{BR}_i(z))^2$$

The terms are squared to achieve an effect similar to the *MostAllocatedFirst* Kubernetes strategy[1], leaving some nodes free to host any incoming heavy job.

Based on this model, we designed a scheduling algorithm for K4.0s. The scheduler addresses two use cases: i) the arrival of a new job, ii) the rescheduling of a job from a node to another. The scheduler uses a greedy strategy that, for each new job, evaluates, for each eligible worker node, what would the objective function score be if the job was assigned to the node. The re-balancing use case computes for each node how much it would benefit from removing each job; selecting the $(node, job)$ pair with the highest increase of objective function score, if the reschedule improves the overall score.

Due to the denominator, accepting a job causes progressively small increases of the acceptance rate. Hence, the scheduler eventually rejects any incoming job not to decrease the score. To overcome this issue, we use as objective function (OF) the product of the weighed sum times the acceptance rate.

## III. PRELIMINARY RESULTS

We tested our K4.0s scheduler comparing it with three possible Kubernetes scheduling configurations[1]:

- *LeastAllocated*: it prefers nodes with more free resources to reduce interferences among jobs.
- *MostAllocated*: it prefers the nodes with less free resource that can host a job (see §II-A).
- *RequestedToCapacityRatio*: it prefers nodes with less percentage of free resources after allocating the job.

In our simulation, all four schedulers filter out worker nodes not eligible to host a job $J$ in the same way; then our scheduler selects the pair $(node, job)$ with the highest OF score, while the others use the above described rules. We extend default Kubernetes' eligibility tests, adding:

- Assurance compliance: $\forall_{k \in Jobs_i \cup \{j\}} A_i^* \geq \theta_k$ where $A_i^*$ is the assurance level of node $i$ if job $j$ were assigned to it, and $\theta_k$ is the assurance threshold of node $k$ based on its criticality level $C_k$
- Real-time compliance: $\neg RT_j \vee RT_i$

The simulation starts with a random number of nodes (between 4 and 10) and jobs (between 300 and 1000). During the experiment, other nodes are randomly added to the cluster and jobs are randomly terminated, freeing the resources allotted to them. Each scheduling algorithm receives the same input sequence, in terms of nodes and jobs.

The plots in Fig. 1 show that co-optimizing the *assurance level* introduced in our model ensures more guarantees to critical jobs without necessarily penalizing the acceptance rate compared to default schedulers.

[1]https://kubernetes.io/docs/reference/scheduling/config/#scheduling-plugins

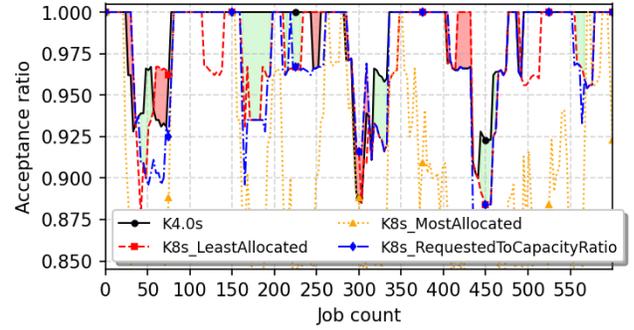

(a) Comparison of acceptance rate

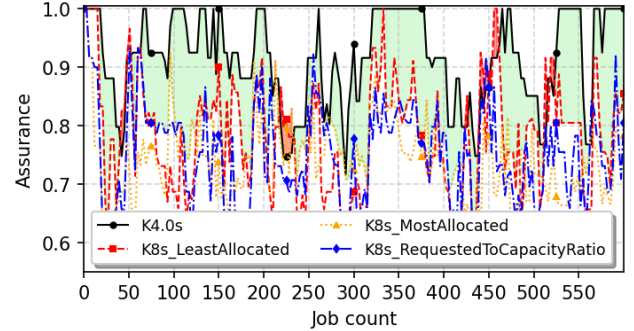

(b) Comparison of average assurance

Fig. 1: Comparison of the objective functions values during the simulation. (Weights for K4.0s: 52.5, 42.5, 5)

## IV. CONCLUSION

These results show the benefit of considering the assurance in the orchestration of mixed criticality environments. However, currently, the assurance is qualitatively computed with a rule-based approach. In our future research, we aim at quantitatively computing and predicting the assurance variation through causal reasoning and Bayesian networks, which perfectly fit with our *what if* scheduling algorithm.

### ACKNOWLEDGMENT

This study was carried out within the MICS (Made in Italy – Circular and Sustainable) Extended Partnership and received funding from the European Union Next-GenerationEU (PIANO NAZIONALE DI RIPRESA E RESILIENZA (PNRR) – MISSIONE 4 COMPONENTE 2, INVESTIMENTO 1.3 – D.D. 1551.11-10-2022, PE00000004). This manuscript reflects only the authors' views and opinions, neither the European Union nor the European Commission can be considered responsible for them.

## REFERENCES

[1] J. Morgan, M. Halton, Y. Qiao, and J. G. Breslin, "Industry 4.0 smart reconfigurable manufacturing machines," *Elsevier Journal of Manufacturing Systems*, 2021.
[2] M. Cinque, D. Cotroneo, L. De Simone, and S. Rosiello, "Virtualizing mixed-criticality systems: A survey on industrial trends and issues," *Elsevier Future Generation Computing Systems*, 2021.
[3] C. Yu, X. Xu, and Y. Lu, "Computer-integrated manufacturing, cyber-physical systems and cloud manufacturing, concepts and relationships," *Elsevier Manufacturing Letters*, 2015.
[4] A. Burns and R. I. Davis, "Mixed criticality systems-a review," *York*, 2022.
[5] V. Struhár, S. S. Craciunas, M. Ashjaei, M. Behnam, and A. V. Papadopoulos, "Hierarchical resource orchestration framework for real-time containers," *ACM Transactions on Embedded Computing Systems*, 2024.
[6] M. Barletta, M. Cinque, L. De Simone, and R. D. Corte, "Criticality-aware monitoring and orchestration for containerized industry 4.0 environments," *ACM Transaction on Embedded Computing Systems*, 2024.
[7] F. Lumpp, F. Fummi, H. D. Patel, and N. Bombieri, "Enabling kubernetes orchestration of mixed-criticality software for autonomous mobile robots," *IEEE Transactions on Robotics*, 2023.